\documentclass[fleqn,10pt]{wlscirep}
\title{Single-pixel imaging with Morlet wavelet correlated random patterns}

\author[1,*]{Krzysztof M. Czajkowski}
\author[1]{Anna Pastuszczak}
\author[1]{Rafał Kotyński}
\affil[1]{University of Warsaw, Faculty of Physics, Warsaw, 02-093, Poland}

\affil[*]{krzysztof.czajkowski@fuw.edu.pl}

\usepackage{placeins}

\begin{abstract}
Single-pixel imaging is an indirect imaging technique which utilizes simplified optical hardware and advanced computational methods. It offers novel solutions for hyper-spectral imaging,  polarimetric imaging, three-dimensional imaging, holographic imaging, optical encryption and imaging through scattering media. The main limitations for its use come from relatively high measurement and reconstruction times.
 In this paper we propose to reduce the required signal acquisition time by using a novel sampling scheme based on a random selection of Morlet wavelets convolved with white noise. While such functions exhibit random properties, they are locally determined by Morlet wavelet parameters. 
 The proposed method is equivalent to random sampling of the properly selected part of the feature space, which maps the measured images accurately both in the spatial and spatial frequency domains. 
  We compare both numerically and experimentally the image quality obtained with our sampling protocol against widely-used sampling with Walsh-Hadamard or noiselet functions. The results show considerable improvement over the former methods, enabling single-pixel imaging at low compression rates on the order of a few percent.
\end{abstract}
\begin{document}

\flushbottom
\maketitle
 
\thispagestyle{empty}
	High resolution detector arrays together with high quality optics constitute the most important parts of any classical camera. 
    Nonetheless, these components, which in some cases tend to be very sophisticated and costly,   are not indispensable elements of imaging systems. Simplifying the optoelectronic hardware of cameras is one of the reasons for the development of indirect imaging techniques. Single-pixel imaging~\cite{IEEE_SPM_5_21_Candes,Baraniuk2008} is a technique which makes use of a single detector, such as a photodiode or photomultiplier, and utilizes spatial and temporal modulation of the optical signal to measure an indirect, compressed and encrypted representation of an image. Currently single-pixel cameras can not compete with the low-cost widely available cameras for the visible wavelength range, however their development offers new possibilities 
    for hyperspectral maging~\cite{Bian:scirep_6_24752,Li:scirep_7_41435}, polarimetric imaging~\cite{OL_37_824_Duran,Soldevila2013}, holographic imaging~\cite{Li:scirep_7_41435,Li2015166,Ramachandran2015110} THz imaging~\cite{Padilla_THz2014},  3D imaging~\cite{Science_Sun2013,Sun:natcommun_7_12010,Li:ao_53_7992} or imaging though scattering media~\cite{Duran:15}, to mention just some applications. Indirect imaging is mostly limited  by the increased time of image acquisition and by the high computational requirements for image reconstruction after the measurement. The branch of mathematics known as compressive sensing~\cite{IEEE_SPM_24_118_Baraniuk,IEEE_SPM_25_14_Romberg,IEEE_SPM_5_21_Candes,Sampling_theory_Eldar}(CS)  brings the tools needed to restore the image from an indirect lower dimensional measurement. The reconstruction problem of the full-dimensional image from such a compressive measurement is  an ambiguous inverse problem consisting in solving an underdetermined  system of linear equations. 

	Images measured by a single-pixel detector are modulated either with structured illumination or using a structured aperture within the detector. As a result, the detector captures a sequence of average intensities of the modulated image. Mathematically, this is a sequence of dot-products of the measured image $X$ with some sampling functions $\psi_i$ which are used for modulation. Usually, the size of measurement is much smaller than the number of pixels of the image at full resolution. This may be seen as a way to capture an encoded and compressed representation of the image which is useful for transmission or storage, and at the same time to deal with the 
    relatively low operation frequencies of current spatial light modulators. For instance, in this work we are using a state-of the art binary spatial light modulator with a maximum resolution of $1024\times 768$ and the maximum frame rate of $22$~kHz. A simple calculation shows that a full measurement with the dimension equal to the number of pixels would take more than half a minute and require $77$GB of memory to store the binary representation of the sampling functions $\psi_i$, which is impractical. Fortunately, the information content of most real-world images is much lower than that theoretically possible to obtain at the same resolution. In other words, most images are well compressible, and an incomplete measurement may carry enough data to obtain an accurate reconstruction of the image at the original resolution.
	
	Most widespread digital image compression methods are adaptive, which means that the compression algorithm is adjusted to the image contents. A different algorithm may be run on various segments of the image, the algorithm may detect constant parts of the image, and after representing the image in a wavelet or other basis, only the highest resulting coefficient are retained. In effect, digital compression algorithms are usually nonlinear, which is difficult to obtain with single-pixel imaging at the stage of image acquisition.
    If some \textit{a priori} information is available on the measured image on top of its compressibility, it makes sense to include this information in the measurement method and to modify the sampling functions accordingly. 
    For instance, a geometrical transformation of the measurement patterns could lead to a nonuniform rate of collecting information from various parts of the image, with the area of interest measured more accurately than the rest~\cite{Phillips:sciadv2017}. Another possibility is to select the subset of measurement patterns that belong to a given basis, for instance consisting of Walsh-Hadamard functions,  not in a random way but rather according to their expected similarity with the image~\cite{Sun:scirep_7_3464}. A dynamic adaptive choice of the sampling patterns may lead to a significant decrease of the size of the measurement~\cite{Huo:ao_55_3356,Assmann:scirep2013}.
    The theory of compressive sensing suggests to use sampling functions which have the smallest coherence with a basis, in which the image has a sparse representation~\cite{Candes:invprob_23_969}. In simple words, most images are well compressible in the wavelet or cosine basis, thus one should use sampling functions which can not be compressed in these bases. Random sampling is a universal choice. For practical reasons, Walsh-Hadamard or noiselet~\cite{Coifman:acha2001,pastuszczak:ao2016} functions are usually used instead, since they are discrete, simple to generate, and a respective representation of the image can be calculated with a fast algorithm, which facilitates image reconstruction.
    In our approach, we develop random sampling functions which sample a specific part of the feature space that we expect to be important for representing a broad class of images.
    
	\section*{Morlet wavelet based nonergodic random sampling}
	Incoherent sampling is 
    based on patterns dissimilar to image contents. On the other hand, a sparse wavelet representation of the image could be also found rapidly by probing the image directly with wavelet functions, if the most probable elements of the wavelet representation are known beforehand. What we propose here, is to combine these two contradictory lines of reasoning into a novel sampling scheme which is both random and based on a wavelet representation at the same time. 
	 We propose to apply a novel kind of sampling, equivalent to random sampling in the feature space. A feature space is built out of vectors, whose elements correspond to specific features of images. Simple features may be  associated with spatial and frequency contents of an image. For instance, a feature space may be constructed using Gabor filters which are defined as Gaussian functions modulated with a linear phase dependence. A two-dimensional Gabor filter $f(x,y)$ has the following form~\cite{Daugman:josaa_2_1160}, 
	\begin{equation}
	f_{x_0,y_0,a,u_0,v_0}(x,y)=N\cdot exp(-\pi[(x-x_0)^2 a^2+(y-y_0)^2 a^2])\times exp(-2\pi i[u_0(x-x_0)+v_0(y-y_0)]),
	\label{eq.gabor2d}
	\end{equation}
	where $N$ is a normalization constant such that $|f|=1$.
	A feature vector is constructed out of a set of dot-products of the image $X$ with Gabor filters $X_{x_0,y_0,a,u_0,v_0}=\langle X,f_{x_0,y_0,a,u_0,v_0}\rangle$ 	where the parameters $x_0,y_0$ are related to probing a certain location of the image, parameter $a$ determines the characteristic scale of the feature, and $u_0,v_0$ select the part of the probed spatial spectrum. Gabor functions allow for probing images in the spatial domain and in the frequency domain at the same time with the highest possible resolution~\cite{Daugman:josaa_2_1160}. In fact, the Fourier representation of the Gabor function is also a Gabor function, and both functions optimize the uncertainty relation for the two-dimensional 
	Fourier transform. In other words, it is not possible to construct narrower probing functions (with smaller variances) in the spatial and frequency domains at the same time. A zero-mean and normalized Gabor filter is known as the Morlet wavelet or Gabor wavelet. In a two-dimensional situation, a Morlet wavelet is equal to
	\begin{equation}
	g_{\sigma,n_p,\theta}(x,y)=Ne^{-\frac{x^2+y^2}{2\sigma^2}}(e^{i(\pi n_p/2\sigma)(x \cos(\theta)+y \sin(\theta))}-\kappa),
	\label{eq.morlet}
	\end{equation}
	where the constants $\kappa$ and $N$ assure that the wavelet function $g$ is normalized $|f|=1$ and has zero mean $\overline{f}=0$. Parameters $\sigma,n_p,\theta$ are related to the size of the Gaussian envelope, number of periods within the envelope, and the orientation of modulation. A feature vector is obtained by convolving the wavelets with the image $X_{x_0,y_0,\sigma,n_p,\theta}= X * g_{\sigma,n_p,\theta}$.
	
\begin{figure}[t]
		\centering
		\includegraphics[width=15cm]{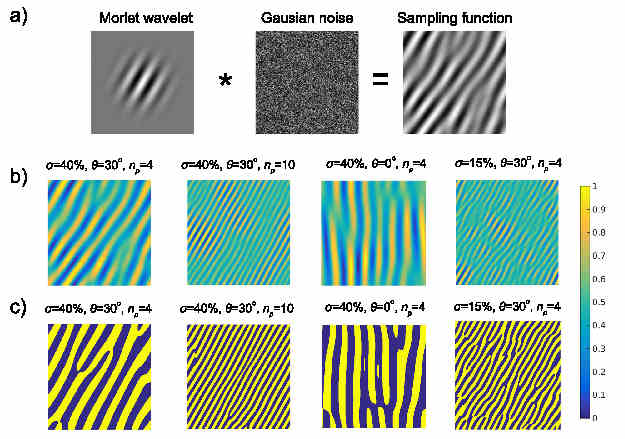}
		\caption{The proposed Morlet wavelet based nonergodic random sampling functions. a)~Schematic of the generation method: the sampling function is calculated by convolving a Morlet wavelet  (see Equation~(\ref{eq.morlet})) with white zero-mean Gaussian noise. b)~Examples of sampling functions with varying parameters ${\sigma,n_p,\theta}$. c)~Examples of binarized sampling functions.}\label{fig.mor}
	\end{figure}

Taking a measurement with a single-pixel detector consists in probing the measured image $X$ with a set of sampling functions $\psi_i$.
The set of measured dot-products $Y_i=\langle X,\psi_i \rangle$ is later used to reconstruct the image $X$.	Now, let us probe the feature space with random functions  $\psi_i$. Since $\langle X*g_{\sigma,n_p,\theta},\psi_i \rangle=\langle X,\psi_i*g_{a,u_0,v_0} \rangle$ instead of probing the feature space, we propose to probe the image $X$ directly with modified sampling functions $\Psi_{\sigma,n_p,\theta}(x,y)=g_{\sigma,n_p,\theta}(x,y)*\psi_i(x,y)$. These sampling functions are obtained as a convolution of Morlet wavelets with realizations of white Gaussian noise. Some examples of sampling functions $\Psi_{\sigma,n_p,\theta}$ and the procedure for their calculation are illustrated in Fig.~\ref{fig.mor}a,b).    

	Many interesting natural phenomena in physics, biology or artificial intelligence arise on the verge of random and deterministic behavior of a system. In this work, the proposed	sampling functions $\Psi_{\sigma,n_p,\theta}$ are calculated by convolving random functions $\psi_i$ and deterministic Morlet wavelets $g_{\sigma,n_p,\theta}$, and they clearly combine random and deterministic properties.  Mathematically, they are zero-mean random matrices with multivariate Gaussian probability density distributions. Each has a distinct power spectrum, dependent on the wavelet used for its generation. As opposed to the wavelets, the sampling functions have a similar random shape at any location, which reflects the property of stationarity. However, their ensemble properties are distinct from the properties of every single realization. This means they do not satisfy the statistical property of ergodicity. This makes them a lot different from the uncorrelated random sampling often used in compressive sensing applications, as well as from the deterministic sampling with noiselet, Walsh-Hadamard or cosine  functions as well as from localized functions  such as wavelets. 

    \begin{figure}[ht]
\centering
	\includegraphics[width=10cm]{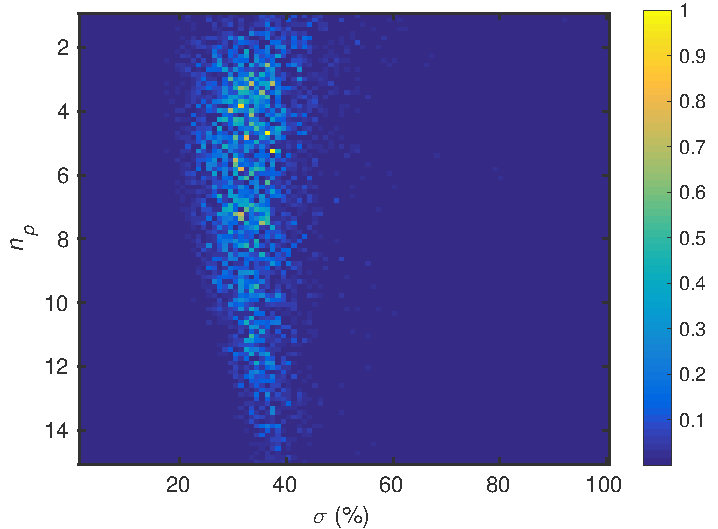}
	\caption{Average decomposition of $49$ test images into a subset of Morlet wavelet-based random sampling functions.}
    \label{fig.optim}
\end{figure}

    	The choice of parameters ${\sigma,n_p,\theta}$ for the random wavelet-based sampling functions has an obvious influence on the quality of a compressive measurement. By decomposing an image database with $49$ images of various content, we found that there exists a common parameter range that may be successfully used to represent most of the images with our sampling functions. This decomposition is not unique and finding an optimal decomposition is challenging from the computational viewpoint. Instead, we have used a simplified approach.  We generated a large number of sampling functions with randomly selected parameters ${\sigma,n_p,\theta}$, placed them into a rectangular matrix, and decomposed every image into these sampling functions by left-multiplying the image by the pseudoinverse of this matrix. The Moore-Penrose pseudoinverse is a generalization of matrix inverse for rectangular and singular matrices,
        it finds application in image reconstruction from compressive measurements~\cite{OE_22_30063_Zhang2014}, and further we also use it as one of the methods for image reconstruction in this paper.
	In this way, we found a typical distribution of coefficients ${\sigma,n_p,\theta}$ required to represent a large variety of real-world images. A graphical representation of the decomposition projected onto the parameter space ${\sigma,n_p}$ is shown in  Fig.~\ref{fig.optim}. As we can see, the interesting part of the feature space spanned with ${\sigma,n_p}$ is easily identified from this plot. A  method with an adaptive choice of sampling functions could be further developed through a more in-depth analysis of how the decomposition varies for different images. However, in our case the range of obtained parameters was similar for the whole image database and the non-adaptive approach taken in this paper is certainly easier to implement, especially in experimental conditions.

 \begin{figure}[ht]
\centering
\includegraphics[width=14cm]{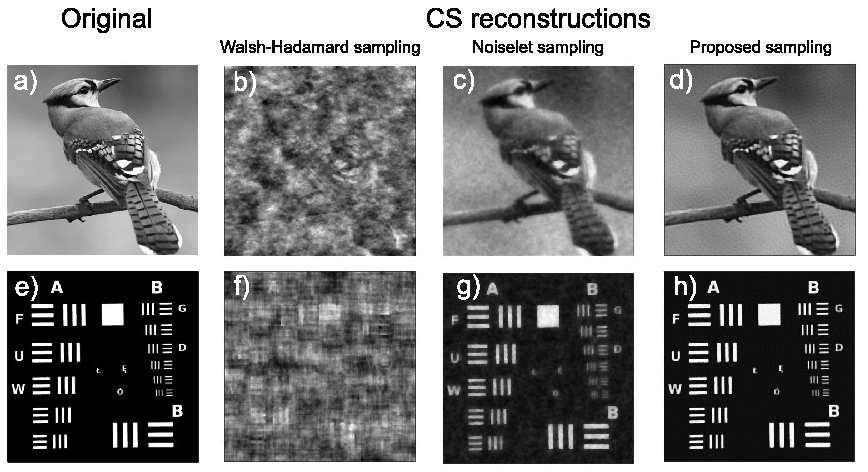}
\caption{CS-based reconstruction of two $512\times 512$ test images from a compressive measurement simulation at the compression rate of 4\%; a,e)~original images; b,f)~reconstruction from a compressive measurement, where sampling was based on Walsh-Hadamard functions;
c,g)~reconstruction from a compressive measurement, where sampling was based on discrete noiselet functions;
d,h)~reconstruction from a compressive measurement, where sampling was based on the proposed Morlet wavelet-based random functions.
}\label{fig.rec1}
\end{figure}

\begin{figure}[ht]
\centering
\includegraphics[width=8cm]{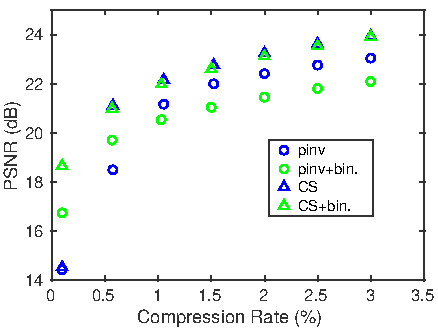}
\caption{Comparison of the image reconstruction quality (PSNR) as a function of compression rate for the fast pseudoinverse-based reconstruction (Pinv) with the slower CS-based reconstruction obtained by minimizing the total variation (TV). PSNR is averaged over a set of ten $512\times 512$ test images, which have been sampled in a numerical simulation using both real and binary Morlet wavelet-based random sampling functions.}\label{fig.err}
\end{figure}

\section*{Results} 
        The practical benefit of using the proposed sampling functions becomes clear from a simple comparison with classical sampling methods based on Walsh-Hadamard or noiselet functions. Figure~\ref{fig.rec1} presents results of a simulated measurement from a single-pixel detector conducted at a low compression rate of~$4\%$ with the use of three different sampling protocols, including the proposed Morlet wavelet-based random functions. The simulation was performed for two~$512\times 512$ images with different properties, such as spatial frequency spectrum, contrast, or richness of details.
        The images have been reconstructed by minimizing the total variation norm (TV), which is one of the basic image reconstruction approaches used in CS. 
        Figures~\ref{fig.rec1}b,f) show the reconstructions from the measurements with Walsh-Hadamard functions, which do not at all resemble the original images. The average quality of the image recovery, measured by the PSNR criterion (see Methods section), for the Walsh-Hadamard sampling is on the level of 14~dB.  This is not surprising at such a low compression rate, although noiselet sampling gives considerably better results (PSNR$\approx22$~dB). On the other hand, the Morlet wavelet-based sampling, also measured at the compression rate of only~$4\%$ allowed us to reconstruct high quality images with the PSNR of 25~dB on average (see Fig.~\ref{fig.rec1}). We would like to emphasize that  these measurements are not adaptive, and the sampling functions are randomly selected from the previously estimated range of values of ${\sigma,n_p,\theta}$, as well as that images shown in Fig.~\ref{fig.rec1}a,e) were not included in the training database.
    Moreover, the quality and resolution of these reconstructions are uniform within the entire image areas and it is not possible to notice any characteristic resolution, orientation or region of interest of the images, which is 
    enhanced at the cost of some other property. We think that this impressive result comes from the efficient sampling of the properly selected part of the feature space which maps the images accurately both in the spatial and spatial frequency domains.   

	Morlet wavelet-based random sampling functions allow to reconstruct images from smaller number of measurements. However, these patterns are neither binary nor orthogonal. We will now discuss the practical consequences of these important limitations and show how to overcome them. 
    In optical single-pixel detectors, light modulation is usually accomplished by using spatial light modulators, such as digital micromirror devices (DMD). The DMDs are capable of displaying binary patterns at a rate of over $20$kHz. Gray-scale modulation could be achieved with time multiplexing or would require using a different spatial modulator, for instance based on liquid crystals. However, both of these approaches offer much lower effective modulation rate.
    Therefore, we have decided to binarize the real-valued Morlet wavelet-based random sampling functions to retain the high measurement performance with a binary DMD. Some examples of binarized sampling functions are shown in Fig.~\ref{fig.mor}c).
    Although binary sampling functions are advantageous for displaying on DMD modulators, they however make the reconstruction of the image more problematic, since the CS algorithms require the basis of sampling functions to be orthogonal. While orthogonalization of a basis of continuous functions is rather straightforward, it cannot be obtained with the binarity constraint placed on the functions. To cope with this problem we use matrices precalculated with the singular value decomposition, which has been explained in the Methods section.
     
   	We have analyzed the influence of the binarization procedure on the quality of the reconstruction of compressively measured images, and we have found it negligible in the case of image recovery with the use of CS optimization method. The reconstruction quality, measured with the peak signal-to-noise (PSNR) criterion  is shown in Fig.~\ref{fig.err} as a function of compression.     
    As an alternative to compressive sensing image reconstruction methods we also make use of a direct recovery from the precalculated pseudoinverse of the measurement matrix. The mathematical details of both kinds of methods are briefly summarized in the Methods section.
    The CS-based recovery offers better quality of the image reconstruction than the pseudoinverse method, especially when binarized sampling patterns are used for image acquisition. However, it can not be obtained in real time. On the other hand, the pseudoinverse-based recovery with the precalculated pseudoinverse matrix requires only the evaluation of a single matrix-vector multiplication, and therefore is very fast. For images of the size of~$256\times 256$ sampled at the compression rate of a few percent, the reconstruction stage is faster than the measurement with the DMD, and may be also realized on-the-fly in parallel with the measurement. 
    
\begin{figure}[ht]
\centering
	\includegraphics[width=11cm]{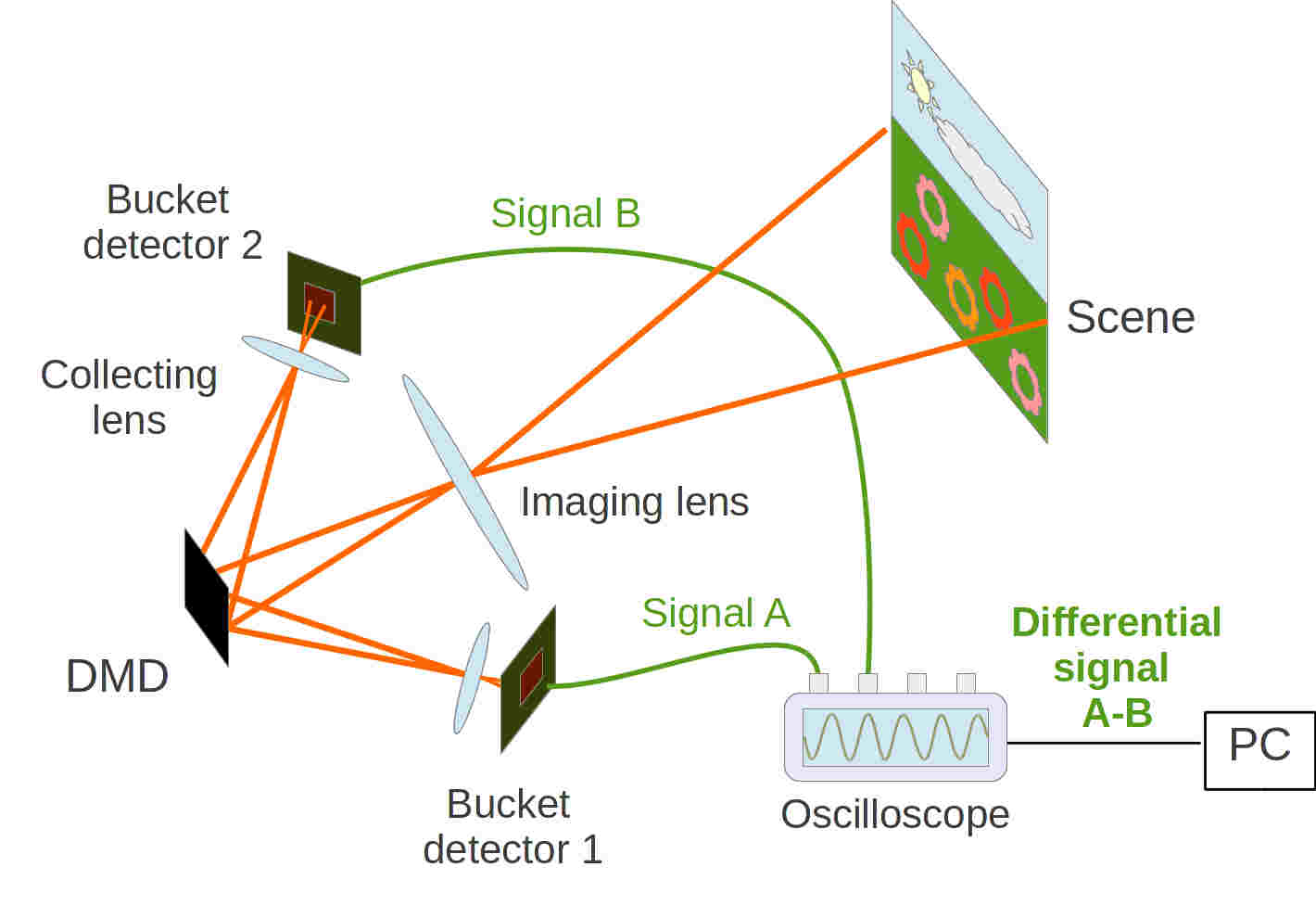}
	\caption{Schematic of the single-pixel detector with differential photodetection. The Si photodiodes measure the light reflected from the DMD mirrors in the on and off states, respectively.}\label{fig.schem}
\end{figure}

Our single-pixel camera set-up shown in Fig.~\ref{fig.schem}  includes a state-of-the-art DMD with $1024\times 768$ pixel resolution and maximum sampling rate of 22~kHz. Signal-to-noise ratio of the measurement is enhanced using the technique of differential photodetection~\cite{Sun:oe2012,Yu:scirep_4_5834,pastuszczak:ao2016}.
The signals measured by two large-area photodiodes are then collected at the rate of 17~MS/s and digitized with 14-bit resolution using a PC oscilloscope.
 \begin{figure}[ht]
\centering
\includegraphics[width=17cm]{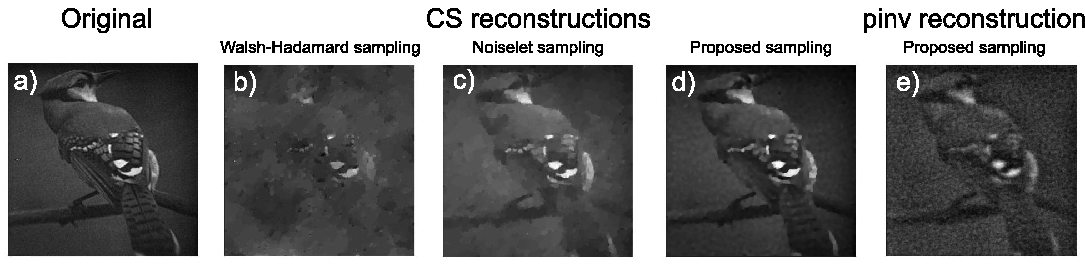}\\
\caption{Experimental comparison of different binary sampling methods. The results are reconstructed at the resolution of $256\times 256$ from a)~a complete measurement, or b,c,d,e)~from a compressive measurement conducted at the compression rate of~$6$\%.  a-d)~have been obtained with a CS algorithm and e) with the pseudoinverse. b)~Walsh-Hadamard sampling; c)~noiselet sampling; d,e)~Morlet wavelet-based random binary sampling.
}\label{fig.exprec}
\end{figure} 
 \begin{figure}[ht]
\centering
\includegraphics[width=8cm]{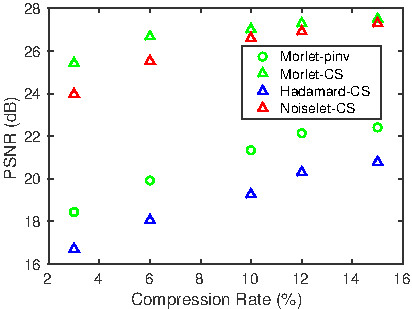}
\caption{Experimental comparison of the image reconstruction quality (PSNR) as a function of compression rate for compressively sensed $256\times 256$ image using Walsh-Hadamard, discrete noiselet, and Morlet wavelet correlated random sampling patterns. For the last case, a fast method of reconstruction based on pseudoinverse (Pinv) is also included, in addition to the CS-based recovery obtained by minimizing the total variation (TV).
}\label{fig.exppsnr}
\end{figure}

A complete measurement at the resolution of $256\times 256$  takes $3$~s and enables us to reconstruct images with a high quality~(see Fig.~\ref{fig.exprec}a), 
restricted only by the imperfections of the experimental set-up and the presence of optical and electronic noise.
Compressive measurements take proportionally less time, however the choice of the sampling protocol is crucial for the feasibility of reconstructing the images with a reasonable quality. For instance, at the compression rate of $6$\%, sampling with a random set of Walsh-Hadamard functions allows to obtain a reconstruction with PSNR on the order of $18$~dB on average~(see Fig.~\ref{fig.exprec}b). At the same time, using the nonergodic Morlet wavelet-based random binary sampling functions leads to the reconstructions with PSNR of over $27$~dB at the same compression rate~(Fig.~\ref{fig.exprec}d), while the PSNR for sampling with discrete noiselet functions is on the order of $25.5$~dB~(Fig.~\ref{fig.exprec}c). Additionally, an image reconstruction obtained with the proposed sampling using pseudoinverse method yields an improved image quality over Walsh-Hadamard sampling with PSNR of $20$~dB. While the noise level is increased in comparison with the iterative CS reconstruction, the pseudoinverse enables rapid image recovery by a single matrix multiplication. 
The sampling scheme proposed in this paper is clearly superior to both other methods also for different compression rates~(see Fig.~\ref{fig.exppsnr}). 




\FloatBarrier
\section*{Discussion}
In this work, we proposed a novel random sampling method for single-pixel imaging. It utilizes nonergodic and stationary Morlet-wavelet-based random patterns that may also be binarized for use with binary spatial light modulators. These sampling functions are obtained as a convolution of Morlet wavelets with realizations of white Gaussian noise. 

The proposed sampling functions have a rich spatial and frequency content. Individually, each is a realization of a multivariate Gaussian noise  with a characteristic feature size, orientation and modulation frequency. Combined together, the sampling functions uniformly probe the feature space spanned over these image features. We have selected a subset of the feature space through the analysis of an image database.
 
We have tested this kind of sampling with a large variety of images, and the proposed method enabled us to reconstruct these images with a good quality at compression rates of just a few percent. Both theoretical and experimental results show that the proposed sampling is a lot better than random Walsh-Hadamard sampling. It is also better than noiselet sampling.

 At such low compression rates it is still possible to use the direct and fast reconstruction method based on the pseudoinverse of the measurement matrix. A direct reconstruction based on a precalculated pseudoinverse matrix may be implemented on-the fly in parallel with image acquisition on a multicore processor. CS-based reconstruction with a better quality requires much longer reconstruction times on the order of seconds.

\section*{Methods}
 \textbf{CS-based image recovery}: We calculate the singular value decomposition of the measurement matrix and following use the total variation~(TV) image recovery method implemented in the  \textit{NESTA}~\cite{Becker:sciam2009} numerical package. When the $k\times n$  measurement  matrix $M$ (with $k<n$, and the compression ratio denoted as  $CR=k/n$) consists of rows with nonorthogonal sampling functions $\Psi_{\sigma,n_p,\theta}(x,y)$, it is first decomposed with the singular-value decomposition (SVD) into a product of small $k\times k$ square orthogonal matrix $U$, diagonal $k\times k$ matrix $D$ and rectangular semiorthogonal complex conjugate transposed $n\times k$ matrix $V$, i.e. $M=U\cdot D\cdot V^{*}$. In effect the TV method operates on orthogonal matrices, as is required to reach convergence. The mathematical model of the measurement $M\cdot X=Y$ (where $X$ is the captured image,  $M$ is the measurement matrix, and $Y$ is the compressive measurement) is replaced with $M'\cdot X=Y'$ (where $M'=V^{*}$, and $Y'=D^{-1}\cdot U^{*}\cdot Y$), with a semiorthogonal matrix $M'$.\\ 
\textbf{Pseudoinverse-based image recovery}: The pseudoinverse of the measurement matrix is calculated through the singular value decomposition $M^{+}=V\cdot D^{-1}\cdot U^{*}$. For the Morlet-based random sampling functions, the measurement matrix and its pseudoinverse have been precalculated before the measurement. In effect, image recovery has been based on a simple matrix-vector multiplication $X\approx M^{+}\cdot Y$, which even for a very large matrix takes a fraction of a second to calculate. Since $[M^{+}\cdot Y]_i=\sum_{j=1}^k{[M^{+}]_{i,j} \cdot Y_j}$, it is possible to calculate this expression on-the-fly during the measurement, as subsequent components $Y_j$ become available.\\
 \textbf{Binarization of the Morlet-based random sampling functions}: The binarization is based on testing the sign of the real-valued functions with the Heaviside step function $\Theta_H$, i.e. $\Psi^{bin}_{\sigma,n_p,\theta}(x,y)=\Theta_H(\Psi_{\sigma,n_p,\theta}(x,y))$. Additionally, a constant function $\Psi^{bin}_1(x,y)\equiv 1$ has been always included as well to measure the mean value of the image.\\
\textbf{Differential photodetection}: the two states of the DMD mirrors direct the reflected light at two different angles. Then two photodiodes measure both $Y_i=\langle X,\Psi^{bin}_i \rangle$ and $ \bar{Y}_i=\langle X,1-\Psi^{bin}_i \rangle$ at the same time. Their difference is used to eliminate the influence of background light and intensity fluctuations of the source from the measurement $Y$.\\
 \textbf{Peak signal to noise ratio}: we use a standard definition of the PSNR for the noisy image $X$ and reference image $R$, $PSNR(X,R) (dB)= 10 \log_{10}(max(R)^2/MSE(X,R))$, where $MSE$ is the mean square error.\\
\textbf{Discrete noiselet 
functions~\cite{Coifman:acha2001}}: Noiselet sampling is a lot less popular than Walsh-Hadamard sampling so we include the definition of discrete noiselets. Let $\mathbf{H}_m$ denote a $m \times m$ Hadamard or noiselet transformation matrix whose rows consist of the basis functions. These matrices may be defined recursively as $\mathbf{H}_{2m}=\mathbf{H}_{2}\otimes \mathbf{H}_{m}$ where $\otimes$ denotes the Kronecker product, $\mathbf{H}_{1}=1$,   and $\mathbf{H}_{2}=\frac{1}{\sqrt{2}} \left[ \begin{array}{cc} 1 & 1 \\ 1 & -1  \end{array} \right]$ for Walsh-Hadamard matrices, and $\mathbf{H}_{2}=\frac{1-i}{2} \left[ \begin{array}{cc} 1 & i \\ i & 1  \end{array} \right]$ for noiselet matrices. Apart from the normalization, Hadamard basis consist of binary  values $\{-1,1\}$, while noiselet basis, depending on $m$, consist of values $\{exp(i p  \pi/4)\}$ with $p=0,2,4,6$ when $m$ is an odd power of $2$, and $p=1, 3, 5 , 7$ when $m$ is an even power of $2$. In the first case the real and imaginary parts of noiselet functions are binary, and in the second the sum and difference of their real and imaginary parts are binary~\cite{pastuszczak:ao2016}.  
Two-dimensional transforms are obtained through the Kronecker product of one-dimensional transforms, i.e. $H_{m\times m}^{2D}=H_m\otimes H_m$.

\section*{Acknowledgements}
The authors acknowledge financial support from the Polish National Science Center (grant No. UMO-2014/15/B/ST7/03107).

 \section*{Author contributions statement}
	K.C. introduced the Morlet wavelet-based sampling functions and conducted numerical analysis. R.K. proposed the idea of nonergodic random sampling.  A.P. constructed the optical set-up and conducted the measurements. All authors reviewed the manuscript.
    
   \section*{Data availability}
   The files generated in this work are available from the corresponding author on reasonable request.

 \section*{Additional information}
\textbf{ Competing interests:} The authors declare that they have no competing interests.


\begin{thebibliography}{10}
\expandafter\ifx\csname url\endcsname\relax
  \def\url#1{\texttt{#1}}\fi
\expandafter\ifx\csname urlprefix\endcsname\relax\def\urlprefix{URL }\fi
\expandafter\ifx\csname doiprefix\endcsname\relax\def\doiprefix{DOI }\fi
\providecommand{\bibinfo}[2]{#2}
\providecommand{\eprint}[2][]{\url{#2}}

\bibitem{IEEE_SPM_5_21_Candes}
\bibinfo{author}{Candes, E.~J.} \& \bibinfo{author}{Wakin, M.~B.}
\newblock \bibinfo{journal}{\bibinfo{title}{An introduction to compressive
  sampling}}.
\newblock {\emph{\JournalTitle{IEEE Sign. Proc. Mag.}}}
  \textbf{\bibinfo{volume}{25}}, \bibinfo{pages}{21--30}
  (\bibinfo{year}{2008}).

\bibitem{Baraniuk2008}
\bibinfo{author}{Duarte, M.} \emph{et~al.}
\newblock \bibinfo{journal}{\bibinfo{title}{Single-pixel imaging via
  compressive sampling}}.
\newblock {\emph{\JournalTitle{IEEE Sign. Process. Mag.}}}
  \textbf{\bibinfo{volume}{25}}, \bibinfo{pages}{83--91}
  (\bibinfo{year}{2008}).

\bibitem{Bian:scirep_6_24752}
\bibinfo{author}{Bian, L.} \emph{et~al.}
\newblock \bibinfo{journal}{\bibinfo{title}{Multispectral imaging using a
  single bucket detector}}.
\newblock {\emph{\JournalTitle{Sci. Rep.}}} \textbf{\bibinfo{volume}{6}},
  \bibinfo{pages}{24752} (\bibinfo{year}{2016}).

\bibitem{Li:scirep_7_41435}
\bibinfo{author}{Li, Z.} \emph{et~al.}
\newblock \bibinfo{journal}{\bibinfo{title}{Efficient single-pixel
  multispectral imaging via non-mechanical spatio-spectral modulation}}.
\newblock {\emph{\JournalTitle{Sci. Rep.}}} \textbf{\bibinfo{volume}{7}},
  \bibinfo{pages}{41435} (\bibinfo{year}{2016}).

\bibitem{OL_37_824_Duran}
\bibinfo{author}{Duran, V.}, \bibinfo{author}{Clemente, P.},
  \bibinfo{author}{Fernandez-Alonso, M.}, \bibinfo{author}{Tajahuerce, E.} \&
  \bibinfo{author}{Lancis, J.}
\newblock \bibinfo{journal}{\bibinfo{title}{Single-pixel polarimetric
  imaging}}.
\newblock {\emph{\JournalTitle{Opt. Lett.}}} \textbf{\bibinfo{volume}{37}},
  \bibinfo{pages}{824} (\bibinfo{year}{2012}).

\bibitem{Soldevila2013}
\bibinfo{author}{Soldevila, F.} \emph{et~al.}
\newblock \bibinfo{journal}{\bibinfo{title}{Single-pixel polarimetric imaging
  spectrometer by compressive sensing}}.
\newblock {\emph{\JournalTitle{Applied Physics B}}}
  \textbf{\bibinfo{volume}{113}}, \bibinfo{pages}{551--558}
  (\bibinfo{year}{2013}).

\bibitem{Li2015166}
\bibinfo{author}{Li, J.}, \bibinfo{author}{Li, H.}, \bibinfo{author}{Li, J.},
  \bibinfo{author}{Pan, Y.} \& \bibinfo{author}{Li, R.}
\newblock \bibinfo{journal}{\bibinfo{title}{{Compressive optical image
  encryption with two-step-only quadrature phase-shifting digital
  holography}}}.
\newblock {\emph{\JournalTitle{Opt. Commun.}}} \textbf{\bibinfo{volume}{344}},
  \bibinfo{pages}{166--171} (\bibinfo{year}{2015}).

\bibitem{Ramachandran2015110}
\bibinfo{author}{Ramachandran, P.}, \bibinfo{author}{Alex, Z.~C.} \&
  \bibinfo{author}{Nelleri, A.}
\newblock \bibinfo{journal}{\bibinfo{title}{{Compressive Fresnel digital
  holography using Fresnelet based sparse representation}}}.
\newblock {\emph{\JournalTitle{Opt. Commun.}}} \textbf{\bibinfo{volume}{340}},
  \bibinfo{pages}{110--115} (\bibinfo{year}{2015}).

\bibitem{Padilla_THz2014}
\bibinfo{author}{Watts, C.~M.} \emph{et~al.}
\newblock \bibinfo{journal}{\bibinfo{title}{{Terahertz compressive imaging with
  metamaterial spatial light modulators}}}.
\newblock {\emph{\JournalTitle{Nature Photon.}}} \textbf{\bibinfo{volume}{8}},
  \bibinfo{pages}{605--609} (\bibinfo{year}{2014}).

\bibitem{Science_Sun2013}
\bibinfo{author}{Sun, B.} \emph{et~al.}
\newblock \bibinfo{journal}{\bibinfo{title}{{3D} computational imaging with
  single-pixel detectors}}.
\newblock {\emph{\JournalTitle{Science}}} \textbf{\bibinfo{volume}{340}},
  \bibinfo{pages}{844} (\bibinfo{year}{2013}).

\bibitem{Sun:natcommun_7_12010}
\bibinfo{author}{Sun, M.} \emph{et~al.}
\newblock \bibinfo{journal}{\bibinfo{title}{Single-pixel three-dimentional
  imaging with time-based depth resolution}}.
\newblock {\emph{\JournalTitle{Nat. Commun.}}} \textbf{\bibinfo{volume}{7}},
  \bibinfo{pages}{12010} (\bibinfo{year}{2016}).

\bibitem{Li:ao_53_7992}
\bibinfo{author}{Li, L.}, \bibinfo{author}{Xiao, W.} \& \bibinfo{author}{Jian,
  W.}
\newblock \bibinfo{journal}{\bibinfo{title}{Three-dimensional imaging
  reconstruction algorithm of gated-viewing laser imaging with compressive
  sensing}}.
\newblock {\emph{\JournalTitle{Appl. Opt.}}} \textbf{\bibinfo{volume}{53}},
  \bibinfo{pages}{7992} (\bibinfo{year}{2014}).

\bibitem{Duran:15}
\bibinfo{author}{Dur{\'a}n, V.} \emph{et~al.}
\newblock \bibinfo{journal}{\bibinfo{title}{{Compressive imaging in scattering
  media}}}.
\newblock {\emph{\JournalTitle{Opt. Express}}} \textbf{\bibinfo{volume}{23}},
  \bibinfo{pages}{14424--14433} (\bibinfo{year}{2015}).

\bibitem{IEEE_SPM_24_118_Baraniuk}
\bibinfo{author}{Baraniuk, R.}
\newblock \bibinfo{journal}{\bibinfo{title}{Compressive sensing}}.
\newblock {\emph{\JournalTitle{IEEE Sign. Proc. Mag.}}}
  \textbf{\bibinfo{volume}{24}}, \bibinfo{pages}{118--121}
  (\bibinfo{year}{2007}).

\bibitem{IEEE_SPM_25_14_Romberg}
\bibinfo{author}{Romberg, J.}
\newblock \bibinfo{journal}{\bibinfo{title}{Imaging via compressive sampling}}.
\newblock {\emph{\JournalTitle{IEEE Sign. Proc. Mag.}}}
  \textbf{\bibinfo{volume}{25}}, \bibinfo{pages}{14--20}
  (\bibinfo{year}{2008}).

\bibitem{Sampling_theory_Eldar}
\bibinfo{author}{Eldar, Y.~C.}
\newblock \emph{\bibinfo{title}{{Sampling theory, Beyond Bandlimted Systems}}}
  (\bibinfo{publisher}{Cambridge Univ. Press}, \bibinfo{year}{2015}).

\bibitem{Phillips:sciadv2017}
\bibinfo{author}{Phillips, D.} \emph{et~al.}
\newblock \bibinfo{journal}{\bibinfo{title}{Adaptive foveated single-pixel
  imaging with dynamic supersampling}}.
\newblock {\emph{\JournalTitle{Sci. Adv.}}} \textbf{\bibinfo{volume}{3}},
  \bibinfo{pages}{e1601782} (\bibinfo{year}{2017}).

\bibitem{Sun:scirep_7_3464}
\bibinfo{author}{Sun, M.~J.}, \bibinfo{author}{Meng, L.~T.},
  \bibinfo{author}{Edgar, M.~P.}, \bibinfo{author}{Padgett, M.~J.} \&
  \bibinfo{author}{Radwell, N.}
\newblock \bibinfo{journal}{\bibinfo{title}{A {Russian} dolls ordering of the
  hadamard basis for compressive single-pixel imaging}}.
\newblock {\emph{\JournalTitle{Sci. Rep.}}} \textbf{\bibinfo{volume}{7}},
  \bibinfo{pages}{3464} (\bibinfo{year}{2017}).

\bibitem{Huo:ao_55_3356}
\bibinfo{author}{Huo, Y.}, \bibinfo{author}{He, H.} \& \bibinfo{author}{Chen,
  F.}
\newblock \bibinfo{journal}{\bibinfo{title}{Compressive adaptive ghost imaging
  via sharing mechanism and fellow relationship}}.
\newblock {\emph{\JournalTitle{Appl. Opt.}}} \textbf{\bibinfo{volume}{55}},
  \bibinfo{pages}{3356} (\bibinfo{year}{2016}).

\bibitem{Assmann:scirep2013}
\bibinfo{author}{Aßmann, M.} \& \bibinfo{author}{Bayer, M.}
\newblock \bibinfo{journal}{\bibinfo{title}{Compressive adaptive computational
  ghost imaging}}.
\newblock {\emph{\JournalTitle{Sci. Rep.}}} \textbf{\bibinfo{volume}{3}},
  \bibinfo{pages}{1545} (\bibinfo{year}{2013}).

\bibitem{Candes:invprob_23_969}
\bibinfo{author}{Candes, E.} \& \bibinfo{author}{Romberg, J.}
\newblock \bibinfo{journal}{\bibinfo{title}{Sparsity and incoherence in
  compressive sampling}}.
\newblock {\emph{\JournalTitle{Inverse Probl.}}} \textbf{\bibinfo{volume}{23}},
  \bibinfo{pages}{969} (\bibinfo{year}{2007}).

\bibitem{Coifman:acha2001}
\bibinfo{author}{Coifman, R.}, \bibinfo{author}{Geshwind, F.} \&
  \bibinfo{author}{Meyer, Y.}
\newblock \bibinfo{journal}{\bibinfo{title}{Noiselets}}.
\newblock {\emph{\JournalTitle{Appl. Comput. Harm. Anal.}}}
  \textbf{\bibinfo{volume}{27--44}}, \bibinfo{pages}{10}
  (\bibinfo{year}{2001}).

\bibitem{pastuszczak:ao2016}
\bibinfo{author}{Pastuszczak, A.}, \bibinfo{author}{Szczygieł, B.},
  \bibinfo{author}{Mikołajczyk, M.} \& \bibinfo{author}{Kotyński, R.}
\newblock \bibinfo{journal}{\bibinfo{title}{Efficient adaptation of
  complex-valued noiselet sensing matrices for compressed single-pixel
  imaging}}.
\newblock {\emph{\JournalTitle{Appl. Opt.}}} \textbf{\bibinfo{volume}{55}},
  \bibinfo{pages}{5141--5148} (\bibinfo{year}{2016}).

\bibitem{Daugman:josaa_2_1160}
\bibinfo{author}{Daugman, J.~G.}
\newblock \bibinfo{journal}{\bibinfo{title}{Uncertainity relation for
  resolution in space, spatial frequency, and orientation optimized by
  two-dimensional visual cortical filters}}.
\newblock {\emph{\JournalTitle{J.Opt.Soc.Am.A}}} \textbf{\bibinfo{volume}{2}},
  \bibinfo{pages}{1160} (\bibinfo{year}{1985}).

\bibitem{OE_22_30063_Zhang2014}
\bibinfo{author}{Zhang, C.}, \bibinfo{author}{Guo, S.}, \bibinfo{author}{Cao,
  J.}, \bibinfo{author}{Guan, J.} \& \bibinfo{author}{Gao, F.}
\newblock \bibinfo{journal}{\bibinfo{title}{Object reconstitution using
  pseudo-inverse for ghost imaging}}.
\newblock {\emph{\JournalTitle{Opt. Express}}} \textbf{\bibinfo{volume}{22}},
  \bibinfo{pages}{30063--30073} (\bibinfo{year}{2014}).

\bibitem{Sun:oe2012}
\bibinfo{author}{Sun, B.}, \bibinfo{author}{Welsh, S.~S.},
  \bibinfo{author}{Edgar, M.~P.}, \bibinfo{author}{Shapiro, J.~H.} \&
  \bibinfo{author}{Padgett, M.~J.}
\newblock \bibinfo{journal}{\bibinfo{title}{Normalized ghost imaging}}.
\newblock {\emph{\JournalTitle{Opt. Express}}} \textbf{\bibinfo{volume}{20}},
  \bibinfo{pages}{16892} (\bibinfo{year}{2012}).

\bibitem{Yu:scirep_4_5834}
\bibinfo{author}{Yu, W.} \emph{et~al.}
\newblock \bibinfo{journal}{\bibinfo{title}{Complementary compressive imaging
  for the telescopic system}}.
\newblock {\emph{\JournalTitle{Sci. Rep.}}} \textbf{\bibinfo{volume}{4}},
  \bibinfo{pages}{5834} (\bibinfo{year}{2014}).

\bibitem{Becker:sciam2009}
\bibinfo{author}{Becker, S.}, \bibinfo{author}{Bobin, J.},  \&
  \bibinfo{author}{Candès, E.~J.}
\newblock \bibinfo{journal}{\bibinfo{title}{Nesta: a fast and accurate
  first-order method for sparse recovery}}.
\newblock {\emph{\JournalTitle{SIAM J. Imaging Sci.}}}
  \textbf{\bibinfo{volume}{4}}, \bibinfo{pages}{1--39} (\bibinfo{year}{2009}).

\end{thebibliography}
\end{document}